\documentclass{aa}
\usepackage{amsmath}
\usepackage{mathtools}
\usepackage{txfonts}
\usepackage{graphicx} 
\usepackage{enumitem}
\usepackage{enumerate}
\begin{document} 
\title{Evolution of spheroidal dust in electrically active sub-stellar atmospheres}
   \author{C. R. Stark
          \inst{1}
          \and
          D. A. Diver\inst{2}
          }
   \institute{Division of Computing and Mathematics, Abertay University, Kydd Building, Dundee DD1 1HG.\\
              \email{c.stark@abertay.ac.uk}
         \and
             SUPA, School of Physics and Astronomy, Kelvin Building, University of Glasgow, Glasgow, G12 8QQ, Scotland, UK.\\ 
\email{declan.diver@glasgow.ac.uk}             
             }
   \date{}
  \abstract
   {Understanding the source of sub-stellar polarimetric observations in the optical and near-infrared is key to characterizing sub-stellar objects and developing potential diagnostics for determining properties of their atmospheres. Differential scattering from a population of aligned, non-spherical dust grains is a potential source of polarization that could be used to determine geometric properties of the dust clouds.} 
   {This paper addresses the problem of the spheroidal growth of dust grains in electrically activated sub-stellar atmospheres. It presents the novel application of a mechanism whereby non-spherical, elongated dust grains can be grown via plasma deposition as a consequence of the surface electric field effects of charged dust grains. }
   {We numerically solve the differential equations governing the spheroidal growth of charged dust grains via plasma deposition as a result of surface electric field effects in order to determine how the dust eccentricity and the dust particle eccentricity distribution function evolve with time. From these results, we determine the effect of spheroidal dust on the observed linear polarization.}
   {Numerical solutions show that $e\approx 0.94$ defines a watershed eccentricity, where the eccentricity of grains with an initial eccentricity less than (greater than) this value decreases (increases) and spherical (spheroidal) growth occurs. This produces a characteristic bimodal eccentricity distribution function yielding a fractional change in the observed linear polarization of up to $\approx0.1$ corresponding to dust grains of maximal eccentricity at wavelengths of $\approx1 \mu$m, consistent with the near infrared observational window.  Order of magnitude calculations indicate that a population of aligned, spheroidal dust grains can produce degrees of polarization $P\approx\mathcal{O}(10^{-2}-1\%)$ consistent with observed polarization signatures.}
 {The results presented here are relevant to the growth of non-spherical, irregularly shaped dust grains of general geometry where non-uniform surface electric field effects of charged dust grains are significant. The model described in this paper may also be applicable to polarization from galactic dust and dust growth in magnetically confined plasmas.}

    \titlerunning{spheroidal plasma dust}
 \maketitle

%

\section{Introduction}
Polarimetric observations in the optical and near-infrared wavelength range are potentially a very useful diagnostic tool for determining the physical properties of sub-stellar environments such as the nature of atmospheric clouds \citep{2001ApJ...561L.123S}. Polarimetric signatures can be the result of a number of phenomena that produce asymmetries in the body's disc, such as inhomogeneous cloud coverage, rotational flattening, differential scattering from a population of aligned, non-spherical dust grains (e.g \citet{2003ApJ...585L.155S}), strong zonal flows \citep{2011ApJ...741...59D}, and the presence of a planet transiting the dwarf's dusty disc (e.g. \citet{2018ApJ...861...41S,miles_paez_2019}). Recently, \citet{2020ApJ...894...42M} argued that the detected linear polarization from Luhman 16A is the result of cloud banding from dynamical atmospheric processes. Therefore, such observations could be used to deduce the physical characteristics of the underlying process producing the signal. This poses a challenge for the modelling and interpretation of such signatures since they could be a superposition of multiple contributions. In order to achieve the true, unambiguous interpretation of the electromagnetic radiation, a thorough understanding of the underlying sources of polarization must be obtained.

Observations of late-M and L dwarfs reveal linear polarization measurements typically smaller than 2 per cent in the I- and J-bands \citep{2017MNRAS.466.3184M,2011ApJ...740....4Z,2005ApJ...621..445Z,2002A&A...396L..35M} and the R-band \citep{2017MNRAS.468.3024M}. In a sample of M, L, and T dwarfs, \citet{2009A&A...502..929G} measured a degree of polarization of $\approx0.31\%$ in one object but failed to detect polarization in the other targets. Moreover, their results failed to correlate with current model predictions for ultra-cool dwarf polarization studies related to flattening-induced or heterogeneous cloud-induced polarization, underlining the complexity of the dynamical atmospheric processes and how they are manifested in a polarimetric signature. For example, \citet{2009A&A...508.1423T} observed polarization variability in 2MASSW J1507476-162738, which could be the result of such complex dynamical processes, including the temporal evolution of cloud coverage in the atmosphere. 

To probe the photopolarimetric characteristics of brown dwarfs, a variety of models have been explored. \citet{2018ApJ...866...28S} have developed a conics-based radiative transfer scheme that calculates the disc-resolved and disc-integrated polarized emission of an oblate object that exhibits patchy clouds. They find that the degree of polarization is sensitive to the oblateness and inclination of the sub-stellar object and that the polarization decreases for large cloud grains. However, their model does not currently consider the effect of non-spherical dust grains on polarization. \citet{2017A&A...607A..42S} have developed a Monte Carlo radiative transfer code for scattered light simulations in exoplanets that can accommodate distributions of gas, clouds, hazes, and circumplanetary material. Amongst other things, they found that Rayleigh scattering by submicron-sized cloud particles maximize the resulting polarimetric signal in the {\it H}-band, whereas the integrated degree of polarization is significantly decreased by micron-sized particles due to forward scattering.

A sub-stellar atmosphere can be considered electrically activated if ionization processes occur creating a population of free electrons and ions. Processes such as thermal ionization~\citep{isabel2015}, electrical discharge events~\citep{helling2013}, cosmic-ray ionization~\citep{rimmer2013,2014ApJ...787L..25R}, turbulence-induced dust-dust collisions~\citep{helling2011b} and Alfv\'{e}n ionization~\citep{stark2013} can create ionized regions where the long-range, collective Coulomb interaction dominates over short-range binary interactions. Thermal ionization is typically significant deep in the atmosphere ($p_{\rm gas}\approx10^{-4} - 10$~bar) and can create ionized regions whose spatial extent and degree of ionization is dependent on the effective temperature, metallicity and $\log{g}$ of the object. For example, as the effective temperature increases from $T_{\rm eff}\approx 1200$~K to $3000$~K, the atmospheric volume fraction that can be considered electrically activated increases from $10^{-3}$ to $1$. The thermal degree of ionization can reach values of $10^{-7}$ to $10^{-4}$~\citep{isabel2015}. The extent the regions can be considered electrically active will depend on the atmospheric temperature distribution within the atmosphere e.g. regions of higher temperature (e.g. see Fig. 3,~\citet{kaspi2015} and Fig. 2,~\citet{lee2016}) may have a greater degree of ionization.
 
For cooler objects thermal ionization may struggle to create a sufficient free electron density to constitute an electrically active atmosphere; however, lower atmospheric temperatures do not necessarily mean that other non-thermal ionization processes cannot occur creating a sufficient density of free electrons (e.g. \citet{2016PPCF...58g4003H}). The possibility exists that the coolest brown dwarfs, such as T and Y dwarfs, could be electrically active to some degree. In the presence of clouds ($p_{\rm gas}\approx10^{-5}-1$~bar) lightning can create regions with degrees of ionization of about $10^{-1}$~\citep{guo2009,beyer2003}, where a single sparking event can occur on length scales of $\approx10^{4}-10^{10}$~m$^{-3}$ \citep{bailey2013}. If there are multiple discharge events the atmospheric volume affected can be enhanced. \citet{gabi2016} estimate that the total number of lightning flashes that can occur in an exoplanet atmosphere during a transit to be of the order of $10^{5}-10^{12}$. Furthermore, Alfv\'{e}n ionization (also known as Critical Velocity Ionization) occurs when a neutral gas collides with a low-density magnetized seed plasma and their relative motion reaches a critical threshold speed. Alfv\'{e}n ionization can create regions with degrees of ionization ranging from $10^{-6}-1$ and is most effective where $p_{\rm gas}\approx10^{-5}-10^{-15}$~bar and where flow speeds are of the order of $1-10$~kms$^{-1}$~\citep{stark2013}. \citet{wilson2019} found that for HD189733b ($T_{\rm eff}=1117$~K) the average ionization fraction at equilibrium as the result of Alfv\'{e}n ionization ranges from $10^{-5}-10^{-8}$ for pressures between $10^{-1}-10^{-3}$~bar.
Cosmic-ray ionization is another non-thermal process that can significantly enhance the degree of ionization for atmospheric pressures $p_{\rm gas}<10^{-2}$~bar. It is expected that cosmic-ray bombardment may create a weakly ionized plasma ( $\gtrsim10^{-8}$) in the upper atmospheres of giant gas planets, where $p_{\rm gas}<10^{-8}$ \citep{rimmer2013}. Further to this, in the presence of a companion or host star photoionization by Lyman continuum radiation can enhance the degree of ionization in the upper atmosphere of sub-stellar objects, creating a shell of ionized material that can interact with the ambient magnetic field potentially leading to chromospheric activity and non-thermal emission \citep{isabel2018}. 

Sub-stellar atmospheres can exhibit radio (e.g. \citet{williams2015,williams2017,route2016}), X-ray (e.g. \citet{audard2007,berger2010}), and optical H-alpha (e.g. \citet{schmidt2007,pineda2016}) emission. Such emission indicates the presence of plasma and the associated high-energy processes in their surrounding envelopes; for example, lightning, synchrotron emission~(e.g. \citet{williams2015}), electron cyclotron maser emission~(e.g. \citet{2012ApJ...760...59N,kao2016}).  The latter mechanism can be used as a precise tool for sub-stellar magnetic field measurements. The observation of such energetic emission is also observed in some L and T brown dwarfs~(e.g. \citet{williams2013,burgasser2013,hall2002}), demonstrating that cooler objects can exhibit plasma behaviour. Beyond direct emission from free electron motion, electron bombardment of neutral species populating the atmosphere (as a result of ionization processes) may excite the neutrals that then relax, producing emission that could be identified in spectra. This excitation may include rotational, vibrational and metastable states, resulting in discrete (and potentially time-dependent) emission lines~\citep{stark2013}. Furthermore, plasma-activity can affect the formation of certain molecules resulting in the weakening of spectral absorption bands such as the 2.7~$\mu$m water line \citep{sorahana2014}. Similarly, \citet{bailey2013} demonstrated that the influence of lightning on the local gas-phase included an increase in the abundance of small carbohydrate molecules like CH and CH$_{2}$ whilst the abundance of CO and CH$_{4}$ decreased.

In an electrically active sub-stellar atmosphere, where ionization processes occur producing free electrons, dust grains can become negatively charged. In such an environment, non-spherical charged dust grains can become aligned to the ambient magnetic field permeating the atmosphere through a number of processes. For example, paramagnetic relaxation of thermally rotating grains \citep{1949PhRv...75.1605D}; diamagnetic alignment \citep{1995JMMM..140.2181U,1998JMMM..177.1455C,1998PASJ...50..149C}; or, via magnetically induced torque of rotating charged dust grains \citep{1996JPhD...29..687D}. In sub-stellar objects it is speculated that their atmospheres are permeated by ambient magnetic fields of the order of $10^{-3}-10^{-1}$~T (e.g. \citet{2009ARA&A..47..333D,2012AJ....143...93R,2009Natur.457..167C,2004ApJ...609L..87S,2011MNRAS.418.2548S}). For example, following \citet{1996JPhD...29..687D} the timescale for the alignment of a dust grain of mass $M$, with charge $Q$ to an external magnetic field, $B$, is of the order of the cyclotron period $P=2\pi M/(QB)$. For $M=4\pi a^{3}\rho/3\approx10^{-15}$~kg (using $\rho\approx10^{3}$~kgm$^{-3}$) and  $Q\approx10^{3}e\approx10^{-16}$~C; yields $P\approx10/B$. For a range of magnetic field strengths of $B\approx 10^{-3}-0.1$~T, the cyclotron period is $P\approx1$~hrs$-1$~mins. Other atmospheric processes occur on the following timescales~\citep{helling2014}: wave propagation ($0.3-3$~s); large-scale convection $20$~min$-3.5$~hrs); gravitational settling ($15$~mins$-8$~months); diffusive eddy mixing ($3$~hrs$-3$~yrs); and, buoyancy oscillations  ($10-10^{3}$~s)~\citep{parent2020}. In comparison, the magnetic field alignment of charged dust grains is possible since it occurs on a shorter timescale than other notable atmospheric processes, with the exception of wave phenomena. However, waves are transitory and spatially localized and so the possibility still exists for intermediate magnetic alignment to occur.

However, in objects where a sufficient magnetic field strength required for alignment is not guaranteed, alignment can still occur via aerodynamical forces acting on the dust particles \citep{2001JAtS...58.2103S,1978JApMe..17.1220P}; or via the collective self-electric field of the charged dust cloud \citep{2007ACP.....7.6161U}. For example, in the former process optical polarimetry observations of a Saharan dust episode indicated the presence of vertically aligned particles in the atmosphere as the result of the collective organization of charged dust cloud particles under the influence of the self-electric field permeating the cloud \citep{2007ACP.....7.6161U}. In the latter process, in the presence of atmospheric fluid flows, axisymmetric particles, such as spheroids, perform end-over-end tumbling motion as the result of the flow interaction. The particles exhibit trivial orientational dynamics becoming aligned with the direction of the instigating flow, damping the tumbling motion as the particles seek the lowest energy configuration state \citep{2016RSPSA.47260226P}. The collective alignment of dust grains via one of the describe processes can result in anisotropic extinction, due to the non-spherical geometry of the dust, yielding an observable polarization signature. In contrast to the other alignment mechanisms, alignment via aerodynamical forces does not require the dust grains to be charged. Temporal variation in the observed polarization signature in these scenarios can be driven by underlying variations in the atmospheric flows, cloud structure or the ambient magnetic field. The process of dust grain alignment requires a mechanism that leads to the formation of non-spherical or elongated dust grains. In an electrically activated  atmosphere, dust grains can grow via ion accretion \citep{2018A&A...611A..91S} where layers of material are electrostatically deposited on the surface of the dust grain. Plasma dust growth occurs when permitted by the local plasma conditions; such activity could straddle conventional fluid atmospheric regions. Critically, the surface electric field of the charged dust determines how the surface material is deposited and the resultant geometry of the growing dust grain \citep{stark2006}.

This paper addresses the problem of elongated growth of dust grains in electrically activated  sub-stellar atmospheres. It presents the novel application of a mechanism whereby elongated, spheroidal dust grains can be grown via plasma deposition as a consequence of surface electric field effects of charged dust grains. The paper is structured as follows: Section~\ref{sec_1} outlines the theoretical model for elongated dust growth in a plasma, summarizing and extending the work of \citet{stark2006}; Section~\ref{sec_2} derives the differential equation governing the elliptical growth of dust grains, determining how the dust eccentricity evolves with time; building on Sect.~\ref{sec_2}, Sect.~\ref{sec_3} quantifies how the dust particle eccentricity distribution function of a population of atmospheric grains evolves with time; Sect.~\ref{sec_4} investigates how the resulting particle eccentricity distribution function affects the observed linear polarization; and Sect.~\ref{sec_5} summarizes the findings of this paper and the consequences for electrically active sub-stellar atmospheres.

\section{Spheroidal dust growth via plasma deposition \label{sec_1}}
We consider a dust grain present in a uniform electron-ion plasma. Due to the greater mobility of the electrons relative to the ions, the dust grain becomes negatively charged, resulting in the formation of an electron-depleted plasma sheath around the grain (e.g. see \citet{2000PSST....9..517B}). As the negative charge on the grain builds up, the number of electrons having the appropriate kinetic energy to overcome the grain potential, striking its surface, decreases. Due to the negative charge on the grain there will now exist a flow of ions towards the grain that are deposited on the grain surface altering its size, mass, charge, and, hence, the potential of the grain. As a result, the number of electrons reaching the grain surface increases, further altering the charge. This charge variation occurs until a particle-flux equilibrium configuration is reached, where  the electron and ion fluxes at the surface of the dust grain are equal, resulting in the dust having a constant negative charge and its surface residing at the floating potential. At this point, the dust grain is surrounded by a plasma sheath with spatial extent of the order of the plasma Debye length, where the potential of the grain is shielded from the plasma. In this steady-state situation
ions will be accelerated towards the grain surface from the plasma by the sheath electric field where they will be deposited ultimately on the grain surface. As ions are accreted onto the surface, a thin layer of material is grown, altering the size and shape of the dust grain. 

Without loss of generality, we will now restrict the problem to a 2D cross-section of a dust grain at the centre of a circular sheath. Consider a uniform circular distribution of ions moving into the sheath from the presheath and heading for deposition on the surface of the  dust grain. The local thickness of the deposited layer depends strongly on the structure of the local electric field. In the case of a spherical grain, with a circular cross-section, the distribution of ions would be deposited uniformly over the surface of the grain if they were to follow the purely radial electric field lines. However, spheroidal dust grains, with an elliptical cross-section, have a non-radial electric field close to the grain and the ions could be deposited on the surface in a non-uniform fashion, leading to growth in a preferred direction. The extent of elongated growth depends on the momentum of the ions: If it is sufficiently low, the ions will closely follow the field lines and will be deposited on the grain surface in a non-uniform fashion, resulting in mass-loading at the grain poles; however, if their momentum is too large, they will not follow the direction of the field lines precisely and resultant elongated growth may not occur.  

There are three distinct dust growth scenarios: non-inertial ions that follow the electric field lines leading to anisotropic growth; purely geometric growth where the spherical input ion flux encounters an elliptical target area; inertial ions moving under the influence of the grain's electric field leading to either spherical or spheroidal growth depending on the ion momentum and the non-uniformity of the grain field. See \citet{stark2006} for further details regarding elliptical growth of grains in the context of supernova remnants.
\begin{figure}
\resizebox{\hsize}{!}{\includegraphics{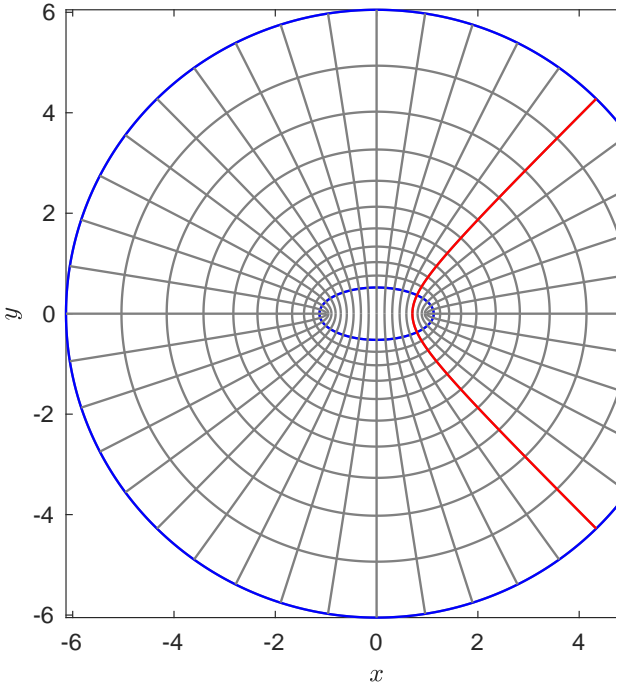}}
 \caption{Elliptical coordinate system $(\xi,\eta)$ consisting of the confocal family of parabolas, $\eta$, and ellipses, $\xi$, depicting the electric field and potential, respectively, of a negatively charged elliptical dust grain in a plasma. The blue ellipse and the circle represent the surface of the dust grain and the plasma sheath, respectively, and the red curve signifies the parabola $\eta=1/\sqrt{2}$ with asymptotes $\theta=\pm\pi/4$.\label{coords}}
\end{figure}
\subsection{Non-inertial ions}
\begin{figure}
\resizebox{\hsize}{!}{\includegraphics{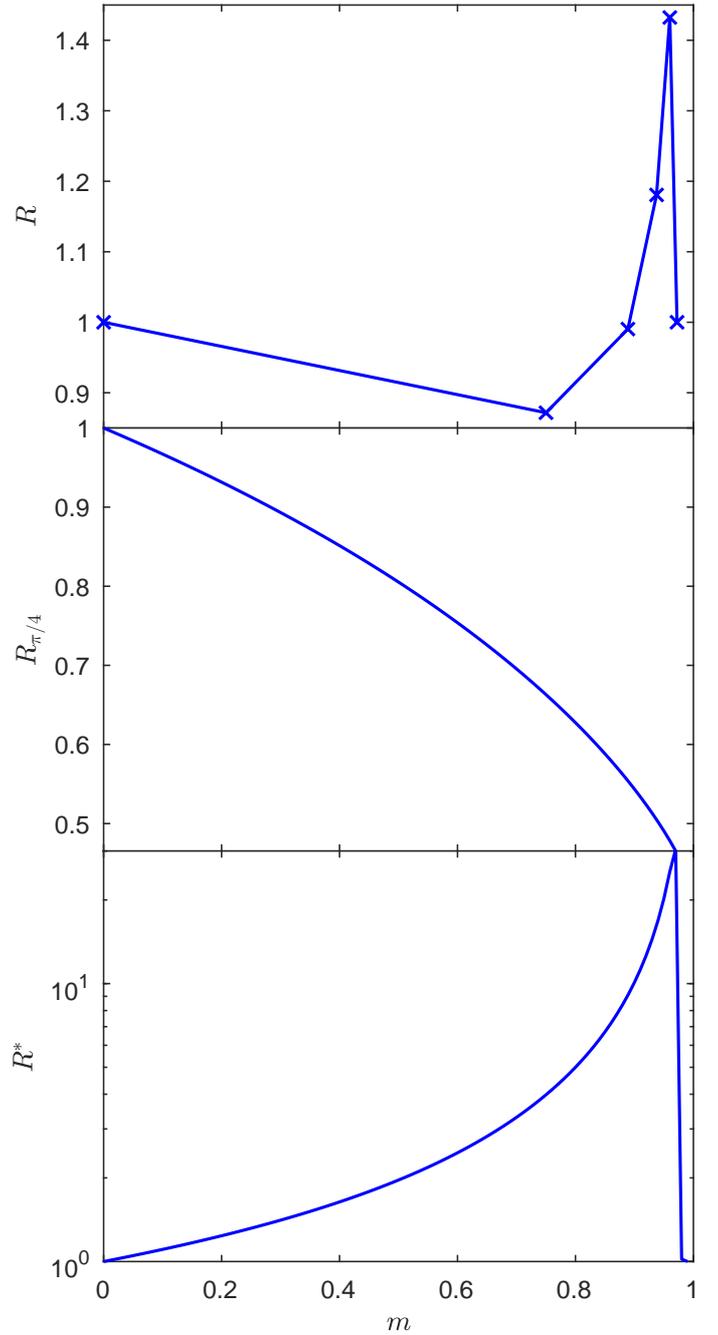}}
 \caption{Measure of elongated grain growth $R$, $R_{\pi/4}$ and $R^{*}$ as a function of $m=e^{2}$. The top plot shows $R$ from numerical simulations for $\alpha\gg1$; the middle plot shows $R_{\pi/4}$ presenting the case $\alpha\ll 1$; and, the bottom plot shows $R^{*}$ the ideal case where deposited ions are non-inertial and elliptical growth always occurs. The $y$-axis of the bottom plot uses a logarithmic scale for ease of readability. \label{r_plot}}
\end{figure}

To help describe the anisotropic mass loading at the grain poles, consider a 2D model cast in elliptical coordinates ($\xi,\eta$) such that the plasma sheath is a circle at the centre of which is an elliptical dust grain (Fig.~\ref{coords}). Therefore, the electric potential (electric field) surrounding the grain can be described by the family of confocal ellipses, $\xi$ (parabolas, $\eta$).  Assuming a uniform circular distribution of ions entering the sheath, the parabola ($\eta=1/\sqrt{2}$) with asymptotes $\theta=\pm\pi/4$, bisects a quadrant of the ring of ions.  If these ions are non-inertial and follow the electric field lines exactly, this ring of material would be deposited on the section of the elliptical grain perimeter that is intersected by the parabola.  The ratio of the two arc lengths in the first quadrant created by the intersection of the ellipse with the parabola $\eta=1/\sqrt{2}$ is given by
\begin{equation}
R^{*}_{0}=\frac{E(\pi/2,m)-E(\theta^{*},m)}{E(\theta^{*},m)}, \label{rstar}
\end{equation}
where
\begin{equation}
\tan{\theta^{*}}=1-m.
\end{equation}
$\theta^{*}$ is the angle subtended by the intersection of the parabola with the grain perimeter at the ellipse centre; $E(\phi,m)$ is the Incomplete Elliptical Integral of the second kind;  $\phi$ is the angle subtended at the centre of the ellipse by the arc; and $m=e^{2}$ is the eccentricity squared.  For elliptical grains ($0<m<1$),~$\theta^{*}<\pi/4$ hence $R^{*}_{0}>1$ and the mass gain of the grains polar region is greater than that of its equatorial region yielding elongated growth. We note that we modify $R^{*}_{0}$ in order to suppress runaway eccentric growth to unphysical values of $m$:
\begin{eqnarray}
 R^{*} &=&(R^{*}_{0}-1)h+1,  \\
 h&=&[\exp{(\sigma(m-m_{\rm max}))}+1]^{-1},
 \end{eqnarray}
where $m_{\rm max}$ is the maximum eccentricity achievable. In this study we define $m_{\rm max}$ to occur when $a=6b$. Figure~\ref{r_plot} (bottom panel) shows $R^{*}$ as a function of the eccentricity squared, $m=e^{2}$. In the case of $R^{*}$, the incoming plasma ions are considered to be non-inertial and so they follow the electric field lines surrounding the dust grain in the sheath exactly. As a consequence, the ions are deposited in a non-uniform, anisotropic fashion on the surface of the dust grain resulting in mass loading at the poles due to the non-radial electric field structure surrounding the dust grain and $R^{*}>1$. The greater the eccentricity of the grain the more pronounced the effect leading to runaway elongated growth until the eccentricity saturates at a maximum value. The $R^{*}$ case is an idealized scenario and unphysical since the incoming ions possess momentum even when the electric potential energy is much greater than the kinetic energy of the ions, presenting an upper-limit to the extent of elongated growth. 
\subsection{Geometric growth}
When the influence of the electric potential on the ion dynamics is very small and $\theta\rightarrow\pi/4$ therefore $R=R_{\pi/4}<1$:
\begin{equation}
R_{\pi/4}=\frac{E(\pi/2,m)-E(\pi/4,m)}{E(\pi/4,m)} \label{rpi4}
\end{equation}
This case is equivalent to classical non-electrostatic gas-phase neutral accretion, where purely geometric growth occurs and the spherical flux of incoming ions encounters an elliptical target area. Figure~\ref{r_plot} (middle panel) shows $R_{\pi/4}$ as a function of the eccentricity squared, $m=e^{2}$. In this case mass loading is greatest at the equator than at the poles of the grain and so $R_{\pi/4}<1$. At $m=0$, $R_{\pi/4}=1$ corresponding to the case of a spherical grain and $R_{\pi/4}<1$ as $m\rightarrow m_{\rm max}$.

\subsection{Inertial ions}
In reality the ions are inertial and the extent of spheroidal growth depends on the dimensional parameter $\alpha=q_{i}\phi_{0}/(m_{i}u_{0}^{2})$, where: $u_{0}$ is the initial speed of the ion (with charge $q_{i}$ and mass $m_{i}$) entering the plasma sheath; and $\phi_{0}$ is the value of the electric potential at the grain surface.  A measure of a grain's growth can be defined as
\begin{equation}
R=\frac{E(\pi/2,m)-E(\theta,m)}{E(\theta,m)},
\end{equation}
where $\theta$ is the angle subtended at the centre of the ellipse by the deposited ion that initial entered the sheath subtending an angle $\pi/4$.  For anisotropic mass loading leading to elongated growth $R > 1$. Numerical simulations conducted to determine the values of $R$ as a function of $m$ and $\alpha$,  found that elongated growth of dust grains occurs only for grains above a certain eccentricity ($m\gtrsim0.88$) when the potential energy of the ions entering the sheath was much greater than their initial kinetic energy $\alpha\gg1$ \citep{stark2006}. Figure~\ref{r_plot} (top panel) shows $R$ as a function of the eccentricity squared, $m=e^{2}$. The case of $R$ exhibits characteristics from both the $R^{*}$ and $R_{\pi/4}$ scenarios. As $m$ increases from $0$, the electric field strength is insufficient to alter the ion trajectories leading to spherical growth $R<1$ (cf. $R_{\pi/4}$ case, when $\alpha\ll1$). This trend is followed until a threshold value of eccentricity is reached $m\approx0.88$ where as $m$ approaches $m_{\rm max}$, the non-radial field is sufficient to alter the ions trajectories leading to elongated growth, $R>1$ (cf. $R^{*}$ case). However, although $R$ exhibits a similar dependence in comparison to the $R^{*}$ case, the increase in $R$ with $m$ is not as severe since the ions can never follow the field lines exactly in the inertial case.

 \section{Eccentricity time evolution  \label{sec_2}}
Using a measure of elongated dust growth such as $R$, a differential equation can be derived that allows the evolution of grain eccentricity as a function of time. Consider an elliptical dust grain with semi-major axis $a$, semi-minor axis $b$ and eccentricity $m=1-b^{2}/a^{2}$.  The temporal evolution of the grains eccentricity $m(t)$, cast in terms of $a(t)$ and $b(t)$ can be written as
\begin{eqnarray}
\frac{dm}{dt}&=&\frac{\partial m}{\partial a} \frac{d a}{d t}+\frac{\partial m}{\partial b} \frac{d b}{d t} \nonumber \\
&=&\frac{2b^{2}}{a^{3}}\frac{da}{dt}
-\frac{2b}{a^{2}}\frac{db}{dt}.
\end{eqnarray}
Writing $\dot{a}=da/dt$ and $\dot{b}=db/dt$ this can be expressed as
\begin{equation}
\frac{dm}{dt}=2(1-m)\frac{\dot{a}}{a}\left( 1-\frac{a}{b}\frac{\dot{b}}{\dot{a}}\right).
\end{equation}
Given $\dot{a}$ and $\dot{b}$ this differential equation describes the variation of a dust grains eccentricity as a function of time $t$.  To encapsulate anisotropic growth we introduce a growth bias factor $\beta$ such that $\dot{a}=\beta\gamma$ and $\dot{b}=\gamma$, where $\gamma$ is the dust growth rate [ms$^{-1}$]. Non-dimensionalizing via $\tau = \gamma t/a_{0}$, $a \rightarrow a/a_{0}$ and $b \rightarrow b/a_{0}$, yields $\dot{a}=\beta$, $\dot{b}=1$ and
\begin{eqnarray}
\frac{\textnormal{d}m}{\textnormal{d}\tau}&=&2(1-m)\frac{\beta}{a}\left(1-\frac{(1-m)^{-1/2}}{\beta}\right), \label{mdot}\\
\frac{\textnormal{d}a}{\textnormal{d}\tau}&=&\beta. \label{adot}
\end{eqnarray}
This has solution,
\begin{eqnarray}
m&=&1-\frac{b^{2}}{a^{2}}=1-\frac{[(1-m_{0})^{1/2}+\tau]^{2}}{a^{2}}, \\
a&=&1+\int \beta\,\textnormal{d}\tau,
\end{eqnarray}
where $m_{0}=1-b^{2}_{0}/a^{2}_{0}$ is the initial eccentricity of a grain; and, $\beta=\beta(m,\tau)$. The growth bias factor can be a complex function of eccentricity and time, leading to non-trivial solutions. In general there are three distinct evolutionary paths that can be followed determined by the parameter, $\delta$:
\begin{equation}
\delta=\frac{a}{b}\frac{\dot{b}}{\dot{a}}=\frac{(1-m)^{-1/2}}{\beta}.
\end{equation}
If $\delta=1$, there is no variation of the eccentricity with time, so $\beta = (1-m)^{-1/2}$, $R=1$ and $\dot{m}=0$. If $\delta<1$, the eccentricity grows with time, $\dot{m}>0$, and is potentially unstable unless capped at a maximum value namely $m=m_{\rm max}$. In this scenario $\beta > (1-m)^{-1/2}$ and $R>1$. An example of this type of behaviour is $\beta=(1-m)^{-1/2}R^{*}$. If $\delta>1$, the eccentricity decays with time, $\dot{m}<0$ and $m\rightarrow 0$. In this scenario $\beta<(1-m)^{-1/2}$ and $R<1$. An example of this case is $\beta=(1-m)^{-1/2}R_{\pi/4}$.

In general, the growth bias factor is of the form $\beta=(1-m)^{-1/2}R$, reflecting that $\beta$ is a complex function of $m$ and that the eccentricity of a grain can either grow or decay in time. 

\begin{figure}
\resizebox{\hsize}{!}{\includegraphics{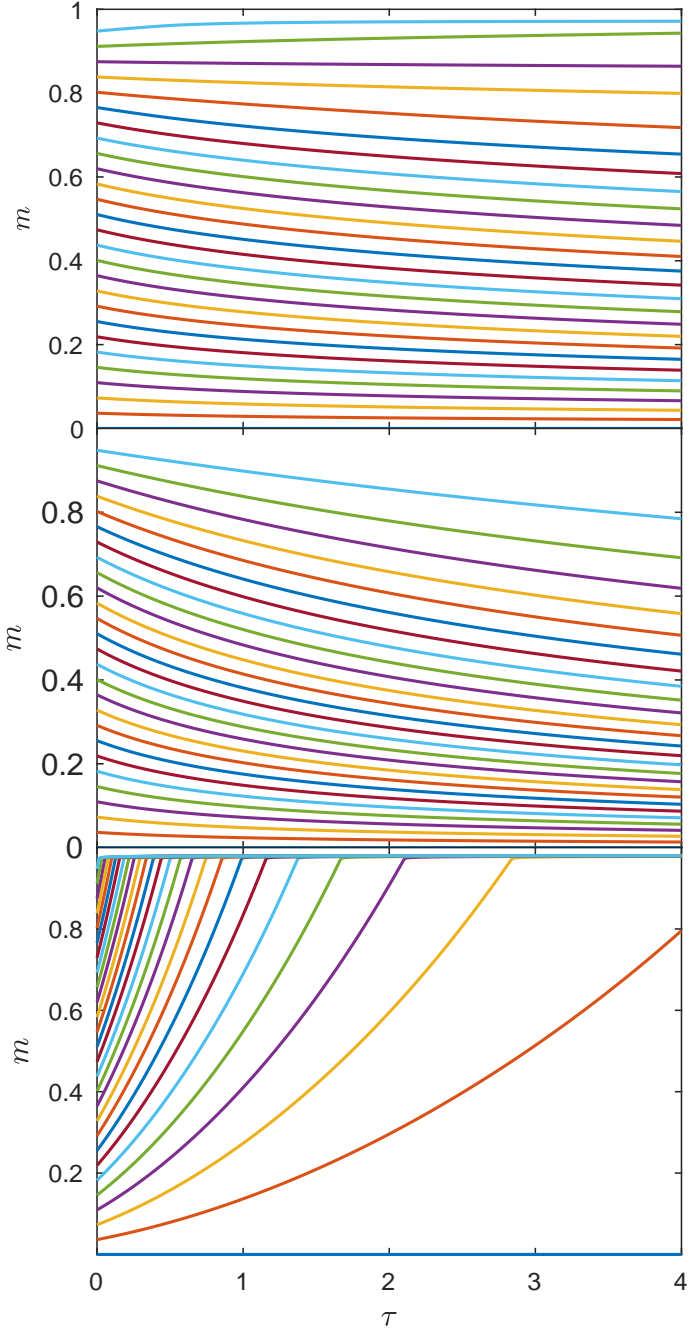}}
 \caption{Evolution of $m=e^{2}$ as a function of time for a range of initial starting eccentricities $m_{0}$ for $R$ (top plot), $R_{\pi/4}$ (middle plot) and $R^{*}$ (bottom plot). In the case of $R$ the point of divergence between grains whose eccentricity increases or decreases with time occurs at the value of $m_{0}$ where $R=1$. In the case $R_{\pi/4}$ the eccentricity of the grains decreases with time for all initial eccentricities; in the case, $R^{*}$ the eccentricity of grains increases with time towards a maximum value for all initial eccentricities.\label{m_plots}}
\end{figure}

Figure~\ref{m_plots} shows the numerical solution of coupled Eqs. (\ref{mdot})-(\ref{adot}) for the cases $\beta=(1-m)^{-1/2}R$ (top panel), $\beta=(1-m)^{-1/2}R_{\pi/4}$ (middle panel) and $\beta=(1-m)^{-1/2}R^{*}$ (bottom panel) using the definitions of $R$, $R_{\pi/4}$ and $R^{*}$ as plotted in Fig.~\ref{r_plot}. At $\tau=0$ a range of initial eccentricities are considered $m_{0}\in[0, m_{\rm max}]$, with $a_{0}=(1-m_{0})^{-1/2}$, and followed as a function of $\tau$.
The evolutionary path for a dust particle is dictated by its instantaneous eccentricity $m$ since $\beta=\beta(m,\tau)$. As ions are accreted onto the surface the eccentricity of the dust grain changes, altering the instantaneous growth bias factor $\beta$ and its onward evolution. In the case of $\beta\propto R_{\pi/4}$ (middle panel), $R_{\pi/4}<1$ for all $\tau$ and so no matter the initial eccentricity, the eccentricity of the dust decreases with time until ultimate spherical growth occurs. When $\beta\propto R^{*}$ (bottom panel), $R^{*}>1$ for all $\tau$ and so for an initial eccentricity (except $m_{0}=0$), the eccentricity of the dust increases with time until it saturates at a maximum value $m_{\rm max}$.  When $\beta\propto R$, $m\approx 0.88$ ($e\approx0.94$) defines a watershed eccentricity: The eccentricity of all grains with an initial eccentricity less than (greater than) this value decreases (increases) and $m\rightarrow0$ ($m\rightarrow m_{\rm max}$).

The variation of eccentricity with times is underpinned by the temporal growth in the semi-major and -minor axes. In this formulation $b=1+\tau$ in non-dimensional form. For a characteristic value for the semi-major axis of $a_{0}\approx10^{-7}$~m at $\tau=4$, the semi-minor axis has grown to~$\approx5\times10^{-7}$~m. For dust growth via plasma deposition \citep{2018A&A...611A..91S}, at $p_{\rm gas}\approx1$~bar, the growth rate is in the range $\gamma\approx10^{-9}-10^{-1}$~ms$^{-1}$ which gives $t=a_{0}\tau/\gamma\approx10^{-6}-10^2$~seconds. In a sub-stellar atmosphere, the growth rate is a function of the ambient properties and so a function of atmospheric pressure.

\section{Particle eccentricity distribution function  \label{sec_3}}
\begin{figure}
\resizebox{\hsize}{!}{\includegraphics{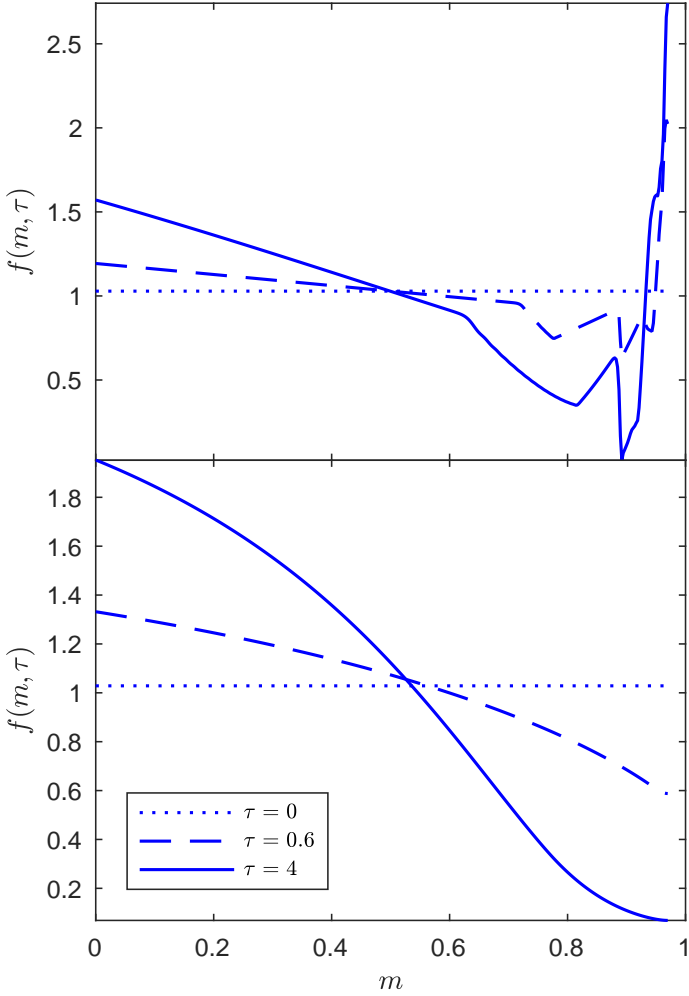}}
 \caption{Eccentricity distribution function as a function of time for $R$ (top plot) and $R_{\pi/4}$ (bottom plot). An initial uniform distribution function (dotted line) is assumed and plotted for an intermediate time (dashed line) and at the end of the simulation (solid line).\label{f_plots}}
\end{figure}

Consider a particle eccentricity distribution function, $g$, as a function of $m$, $b$ and $\tau$, $g=g(m,b,\tau)$. The flux of spheroidal particles as they evolve in time is $(\dot{m}g,\dot{b}g)$; therefore, for a population of spheroidal dust grains with a distribution of eccentricities, the continuity equation governing their evolution is given by
\begin{eqnarray}
\frac{\partial g}{\partial \tau}+\frac{\partial}{\partial m}\left(g\frac{\textnormal{d}m}{\textnormal{d}\tau}\right)
+\frac{\partial}{\partial b}\left(g\frac{\textnormal{d}b}{\textnormal{d}\tau}\right)&=&0 \nonumber \\
\frac{\partial g}{\partial \tau}+\frac{\partial}{\partial m}\left(g\frac{\textnormal{d}m}{\textnormal{d}\tau}\right)
+\frac{\partial g}{\partial b}&=&0. \nonumber 
\end{eqnarray}
For simplicity, to focus on $m$ evolution only we assume
\begin{equation}
g(m,b) = f(m)\delta(b-b_{0})N_{0},
\end{equation}
where $N_{0}$ is the total number of dust particles in the distribution. Hence,
\begin{equation}
\frac{\partial g}{\partial b}=f(m)N_{0}\frac{\partial}{\partial b}[\delta(b-b_{0})]=0
\end{equation}
since
\begin{equation}
\delta^{\prime}(x)=-\frac{\delta(x)}{x}=0.
\end{equation}
Therefore,
\begin{equation}
\frac{\partial f}{\partial \tau}=-\frac{\partial}{\partial m}\left(f\frac{dm}{d\tau}\right).
\end{equation}
This equation is numerically solved using the FDTD-RK4 method. Figure~\ref{f_plots} shows the evolution of $f$ as a function of $\tau$ for an initial flat distribution function,
\begin{equation}
f(m,\tau=0) = \frac{1}{m_{\rm max}},~~~~~~~~\textnormal{for}~m\in[0,m_{\rm max}]. 
\end{equation}
Assuming an initial flat distribution function in $m$ allows the departure of $f$ as a function of $\tau$ and $m$ to be easily quantified without loss of generality.

\begin{figure}
\resizebox{\hsize}{!}{\includegraphics{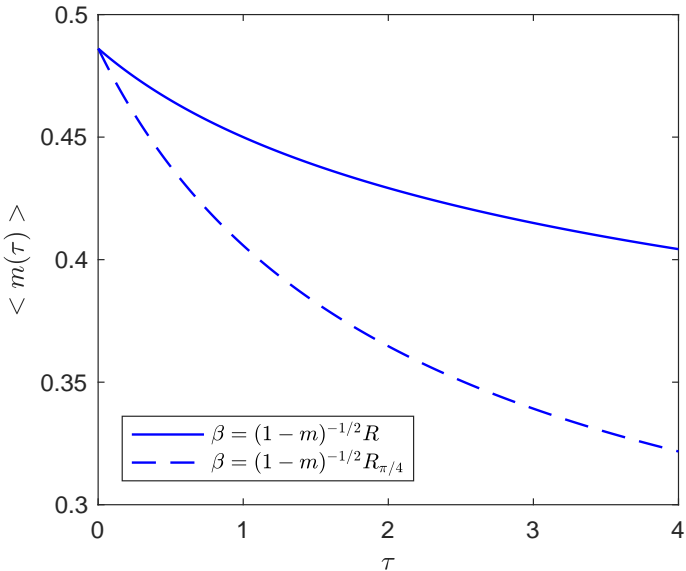}}
 \caption{Mean eccentricity as a function of time calculated from the evolving distribution functions for case $R$ (solid line) and $R_{\pi/4}$ (dashed line).\label{mean_plots}}
\end{figure}

Figure~\ref{f_plots} shows the particle eccentricity distribution function as a function of $\tau$ for an initial flat distribution function and $\beta\propto R$ (top panel) and $\beta\propto R_{\pi/4}$ (bottom panel) taken from the top and bottom plots of Fig.~\ref{m_plots}, respectively.  In both cases dust particles migrate to lower values of eccentricity with time leaving a depleted population at higher eccentricities. However, in the $R$ case dust grains with an eccentricity greater than the watershed value move to higher eccentricities until a maximum value is obtained, resulting a peak in the distribution at $m_{\rm max}$. As a result, in a closed system where the initial distribution of dust particles is not replenished, particles with eccentricities around the watershed value are gradually excavated until the distribution function is exhausted and $f=0$. This produces a bimodal distribution function separated by a region in eccentricity-space devoid of dust particles. As a consequence the mean eccentricity $\langle m\rangle$ as a function of $\tau$ (see Fig.~\ref{mean_plots}) deviates from what is expected classically: $\langle m \rangle$ decreases more slowly as a consequence of the large eccentricity population that persists in the $R$ case.
%
%
%
%
%
\section{Consequences for polarization  \label{sec_4}}
\begin{figure}
\resizebox{\hsize}{!}{\includegraphics{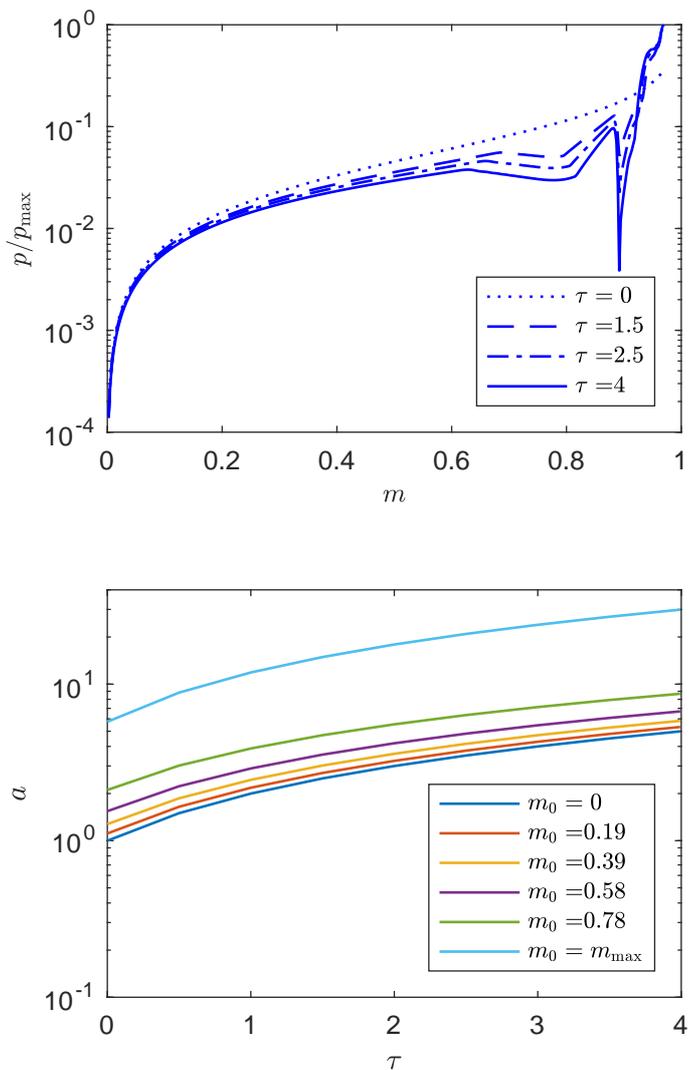}}
 \caption{Top plot: Fractional change in the linear polarization as a function of the particle eccentricity distribution function, $f$, (see top plot Fig.~\ref{f_plots}) and time $\tau$. Bottom plot: Evolution of the semi-major axis $a$ (in dimensionless units) with time $\tau$ for a range of initial eccentricities, $m_{0}$. \label{p_plots_2}}
\end{figure}

If non-spherical dust grains can be collectively aligned via a magnetic field~(e.g.~\citet{1996JPhD...29..687D,1998JMMM..177.1455C,1949PhRv...75.1605D}), aerodynamical forces~(e.g. \citet{2001JAtS...58.2103S,1978JApMe..17.1220P,2016RSPSA.47260226P}) or electric field effects~(e.g. \citet{2007ACP.....7.6161U}), the resulting array of oriented dust grains will preferentially scatter and absorb electromagnetic radiation traversing the atmosphere. There are a number of mechanisms that can align the dust grains in order to achieve a detectable signature. The particle eccentricity distribution function of an evolving population of spheroidal dust grains (such as in Fig.~\ref{f_plots}, top plot) in a sub-stellar atmosphere can give an indication of the effects of the dust on the observed linear polarization. The linear polarization, $p$, as a result of non-uniform extinction due to asymmetric dust is defined as,
\begin{equation}
p = A_{\parallel}-A_{\perp},
\end{equation}
where $A$ is the extinction, in terms of apparent magnitude, when the $\vec{E}$-vector of the EM radiation is parallel  ($\parallel$) and perpendicular ($\perp$) to the long axis of the dust grain \citep{whittet2003}. We can write the extinction as
\begin{eqnarray}
A&=&-2.5\log{(I / I_{0})} \nonumber \\
&=&2.5\log{(\exp{(-\tau)})} \nonumber \\
&=&1.086N_{d}C_{\rm ext}, 
\end{eqnarray}
where
\begin{equation}
\tau=\int n_{d}\textnormal{d}l\cdot C_{\rm ext} =N_{d}C_{\rm ext}
\end{equation}
is the optical depth, $N_{d}$ is the column density of particles; $n_{d}$ is the number density per unit volume; and $C_{\rm ext}$ is the extinction cross section of the dust population since we assume that the material composition of the grains is similar. Therefore, the linear polarization can be written as
\begin{eqnarray}
p&=&1.086N_{d}\sigma(Q_{\parallel}-Q_{\perp}), \nonumber \\
&=&1.086N_{d}\pi a b\left(\frac{C_{\rm ext}}{\pi a^{2}}-\frac{C_{\rm ext}}{\pi b^{2}}\right), \nonumber \\
&=&1.086N_{d}C_{0}ma/b, \label{pole}
\end{eqnarray}
where we have assumed that $Q$ (the extinction efficiency factor) is controlled by the geometric cross-section and not $C_{\rm ext}$. This expression encapsulates both the eccentricity, the scale size and the relative abundance of evolving dust particles. Therefore, we can define a measure of the fractional change in the linear polarization as a function of the particle eccentricity distribution function:
\begin{equation}
\frac{p}{p_{\rm max}}=\left(\frac{f}{f_{\rm max}}\right)\left(\frac{m}{m_{\rm max}}\right)\left(\frac{a}{a_{\rm max}}\right)\left(\frac{b_{\rm max}}{b}\right), \label{p_fract}
\end{equation}
where we have normalized $p$ to the maximum linear polarization $p_{\rm max}$ obtained from the simulations presented in the top plot of Fig.~\ref{f_plots} for the case $\beta=R$. Fig.~\ref{p_plots_2} shows Eq.~\ref{p_fract} plotted as a function of $f$, $m$, $a$, and $b$: The top plot exhibits the fractional change in the linear polarization as the distribution function evolves in time $\tau$; the bottom plot shows the evolution of the semi-major axis $a$ with time $\tau$ for a range of initial eccentricities, $m_{0}$, showing the range of expected grain length scales as the grains evolve. There is a significant fractional change in the linear polarization for $m>0.6$ and in particular at $m\approx0.9$ ($p/p_{\rm max}\approx10^{-2}-10^{-3}$) and $m\approx m_{\rm max}$ ($p/p_{\rm max}\approx0.1$). This is a result of the development of a bimodal distribution of eccentricities as particles with $m\lesssim0.88$ migrate to lower eccentricities and particles with $m\gtrsim0.88$ migrating to higher eccentricities until saturation occurs at $m_{\rm max}$. For monochromatic illumination, Fig.~\ref{p_plots_2} can give an indication of the resonant wavelength at which the dust particle's polarization will be visible. The fractional change in linear polarization corresponds to particles with semi-major axes of $a\approx1-10$ in dimensionless units (Fig.~\ref{p_plots_2}, lower plot). Dust particles with a length scale comparable to the wavelength of the interacting radiation will resonate most effectively with the radiation; therefore, the polarization signature will be observed at $\lambda\approx a\approx1-10$ in dimensionless units. For the case $a_{0}=10^{-7}$~m,  for $\lambda\approx1~\mu$m (i.e. corresponding to the line $a = 10$ on the bottom plot of Fig.~\ref{p_plots_2}) there is no polarization signal until $\tau=0.6$ when dust grains with $m_{0}\gtrsim 0.88$ grow into the observational window and resonate at $\lambda\approx1~\mu$m producing a polarization signature at $m\gtrsim 0.88$. For the case $a_{0}=10^{-6}$~m,  for $\lambda\approx3~\mu$m (i.e. corresponding to the line $a = 3$ on the bottom plot of Fig.~\ref{p_plots_2}) initially there is an observable polarization signature at $m_{0}\approx0.8$, but as the grains grow in size this signal decreases and disappears when $\tau\approx2.5$ as the polarization signal shifts to longer wavelengths. In such scenarios, the population of particles will grow outwith the near-infrared wavelength range to longer wavelengths resulting in the characteristic polarization signature being observed at longer wavelengths. For monochromatic illumination, the evolution of dust grains will result in a transient polarization signature at a given wavelength. In an environment where there is a constant replenishment of dust grains with a range of eccentricities and sizes, creating a dynamical equilibrium, a persistent polarization signature at a range of wavelengths will occur. Such replenishment can come from the fracture of larger grains that will seed the growth of a bimodal population of dust particles. 

To give an order of magnitude estimation of the expected degree of polarization as a consequence of a population of aligned spheroidal dust grains, we recast Eq.~(\ref{pole}) as
\begin{equation}
p\approx10^{-15}\zeta L\kappa_{d}n_{d}m(1-m)^{-1/2},
\end{equation} 
using $C_{0}=\kappa_{d}\rho_{d}/n_{d}$, $\rho_{d}=n_{d}m_{d}$, $N_{d}=n_{d}L$ and $m_{d}=\frac{4}{3}\pi ab^{2}\rho_{m}$, where: $\rho_{d}$ is the dust mass density; $m_{d}$ is the mass of a dust grain of material density $\rho_{m}\approx10^{3}$\,kgm$^{-3}$; $\kappa_{d}$ is the dust opacity; $L$ is a characteristic atmospheric length scale reflecting the spatial extent of an atmospheric region containing a population of aligned spheroidal dust grains; and, $\zeta\in[0,1]$ is free parameter that quantifies that not all grains will be perfectly aligned, nor will the entire atmospheric region will be populated by spheroidal dust grains. We note that this expression does not take into account the effects of the atmospheric gas-plasma. The degree of polarization, $P$, is related to the linear polarization, $p$, if the polarization is sufficiently small, via the expression $P\approx46.05p$ \citep{whittet2003}; therefore, 
\begin{equation}
P\approx10^{-14}\zeta L\kappa_{d}n_{d}m(1-m)^{-1/2}. \label{big_p}
\end{equation}
For typical values, $n_{d}\approx10^{10}$\,m$^{-3}$ (e.g. \citet{helling2014}), $\kappa_{d}\approx10^{-4}$\,m$^{2}$kg$^{-1}$ (e.g. \citet{helling2008b,lee2016}, $\zeta\approx10^{-1}$, $L\approx10^{7}$\,m (e.g.~\citet{isabel2018}), $a\approx10^{-6}$~m and $m\approx0.9$, the degree of polarization , $P$, is of the order $\approx10^{-2}\%$ at $\lambda\approx1$\,$\mu$m. This is consistent with degrees of polarization observed from sub-stellar objects. This is a first approximation calculation and serves the purpose of demonstrating that feasible values for the degree of polarization are achievable; however, we note that the implicated parameters can have a range of values that affect the degree of polarization. For example, altering the dust number density or the opacity by two orders of magnitude can yield a degree of polarization $P\approx\mathcal{O}(1\%)$. In principle, Eq.~(\ref{big_p}) could be used to diagnose properties of the dust clouds, such as the dust density $n_{d}$ or the average dust eccentricity, via polarimetric observations if the other implicated parameters can be independently determined. To further quantify the effects of spheroidal growth on the polarization signatures of sub-stellar objects it would be instructive to combine the model presented here with a theoretical atmospheric model and incorporate a detailed radiative transfer calculation. This is being considered by the authors.

%
%
%
%
%
%
\section{Discussion  \label{sec_5}}
This paper has addressed the problem of non-spherical, elongated growth of dust grains in electrically activated  sub-stellar atmospheres and their contribution towards the generation of a polarimetric signature. It has presented the novel application of a mechanism whereby elongated, spheroidal dust grains can be grown via plasma deposition as a consequence of the surface electric field effects of charged dust grains. We have derived the coupled set of differential equations governing the evolution in time of the eccentricity squared $m=e^{2}$ and the scale size of a dust grain as well as the differential equation describing the dust particle eccentricity distribution function of a population of dust grains as a function time. Results from numerical solutions show that $m\approx 0.88$ ($e\approx0.94$) defines a watershed eccentricity, where the eccentricity of grains with an initial eccentricity less than (greater than) this value decreases (increases) and $m\rightarrow0$ ($m\rightarrow m_{\rm max}$). This produces a characteristic bimodal eccentricity distribution function yielding a fractional change in the observed linear polarization of up to $\approx0.1$ corresponding to dust grains of $m\approx m_{\rm max}$ at wavelengths of $\approx1 \mu$m, consistent withe the near infrared observational window. Order of magnitude calculations indicate that a population of aligned, spheroidal dust grains can produce degrees of polarization $P\approx\mathcal{O}(10^{-2}-1\%)$ consistent with observed polarization signatures.

The collective alignment of charged spheroidal dust grains may be a contributing process resulting in the observed polarization signatures and photometric variability apparent in L dwarfs. In the case of an L dwarf exhibiting variability but no polarization, this is consistent with the result of patchy cloud coverage consisting of spherical dust particles; whereas, an L dwarf exhibiting no variability but polarization could be the result, for example, of homogeneous cloud coverage consisting of aligned, spheroidal dust particles. An L dwarf exhibiting both variability and polarization is consistent with the result of patchy cloud coverage consisting of a population of spheroidal dust particles. The absence of polarization could be due to the inability to grow non-spherical dust grains, or to collectively align the dust grains in a preferred direction.  Non-spherical growth could be disrupted due to the absence of significant ionization triggering the ionic growth process; whereas, the lack of collective grain alignment may be the result of insufficient cloud charging, unfavourable atmospheric flows or weak ambient magnetic fields. Temporal variations in these contributing factors would lead to variability in the resulting polarization signature occurring on the respective natural timescales of evolution. In order to achieve the true, unambiguous interpretation of observed polarization signatures from sub-stellar objects, a thorough investigation and understanding of the possible underlying sources of polarization must be obtained such as inhomogeneous cloud coverage (e.g.~\citet{2020ApJ...894...42M}), rotational induced oblateness (e.g.~\citet{2010ApJ...722L.142S}), strong zonal flows (e.g.~\citet{2011ApJ...741...59D}), the presence of a planet transiting the dwarf's dusty disc (e.g.~\citet{miles_paez_2019}) and the differential scattering from a population of aligned, non-spherical dust grains (e.g.~\citet{2003ApJ...585L.155S}). This poses a challenge for the modelling and interpretation of polarization signatures since they could be a superposition of multiple contributions.

Although this paper has discussed the evolution of spheroidal dust, the results are also of importance to non-spherical dust grains of any geometry. The process discussed here quantifies the effects of non-uniform surface electric fields of charged dust grains resulting from tip geometry effects. Therefore, non-spherical, irregular growth of charged dust grains may occur if non-uniform surface charging of dust grains occurs. The resulting polarization signature will express the differential extinction as a result of the inhomogeneous extinction efficiency factor of the non-spherical dust particles. This paper has focussed on dust growth in the plasma deposition regime where the plasma electron temperature is sufficiently low. In the plasma sputtering regime, where the electron temperature is sufficiently high, the incoming ions may collisionally eject surface material \citep{2018A&A...611A..91S}. This could lead to the spheroidal grains being `etched' back to spherical grains in a reverse evolutionary process, removing high $m$ grains in local regions where electron temperatures permit sputtering.

The presence of spheroidal dust has further consequences beyond its effect on polarization. For a given volume, the surface area of a spheroid is greater than that of a sphere and so the atmospheric gas sees a greater effective surface area for surface chemistry (e.g. \citet{stark2014,stark2015}). Sparking between charged dust grains \citep{helling2011a} can electrically activate the environment allowing chemical reactions to occur that would not be achievable through thermal excitation alone \citep{bailey2013}. Due to the tip-effect of charged objects, spheroidal dust grains will have enhanced electric field strengths at their poles enhancing inter-grain electrical discharges. 

A population of non-spherical dust grains in sub-stellar clouds, as well as potentially yielding a polarization signature, will affect the emergent flux at different wavelengths in a similar way to the growth of spherical dust grains consistent with contemporary atmospheric models (e.g.~\citet{tsuji2002,ackerman2001,witte2009,2012RSPTA.370.2765A,2017AA...608A..70J,helling2008b,powell2018,morley2012,lee2016,lines2018}). This will be dependent on the evolutionary timescale for the dust particle size distribution and the effect of dust growth on the depletion of the local ambient gas-plasma mixture in the atmosphere. In the case of non-spherical dust grains grown via plasma deposition the growth timescales can be faster in comparison to neutral gas-phase surface chemistry. This can lead to the alteration of the local particle size distribution function, in electrically active regions, on a timescale consistent with observed variations in cloud cover, possibly giving a source of atmospheric variability~\citep{2018A&A...611A..91S}.

The results are presented in the context of electrically activated  sub-stellar atmospheres but the mechanism has applications in other areas where a gas is electrically activated ; for example, in the interstellar medium elongated dust grains are believed to be responsible for the observed galactic polarization signatures (e.g. \citet{2019ApJ...883..122L}) potentially opening an diagnostic window to the local environmental conditions. Outwith astrophysics, in magnetically confined fusion, material sputtered from the vessel walls has an elongated geometry (e.g. \citet{arnas:hal-01499487}) and can disrupt the optimum operation of the device when it migrates through the plasma. Understanding the growth and evolution of such dust can help mitigate disruption and enhance operation.
\begin{acknowledgements}
The authors are grateful to the anonymous referee for constructive comments and suggestions that have improved this paper. CRS is grateful for funding from the Royal Society via grant number RG160840 and from the Carnegie Trust for the Universities of Scotland via research incentive grant number RIG007788. DAD is grateful for funding from EPSRC via grant number EP/N018117/1. 
 \end{acknowledgements}
\bibliographystyle{aa}
\bibliography{spheroid_pap}
\end{document}